\documentclass[aps,prl,reprint,groupedaddress,amsmath,amssymb,superscriptaddress]{revtex4-2}
\usepackage{graphicx}
\usepackage{float}
\usepackage{multirow}
\usepackage{natbib,twoopt}
\usepackage[breaklinks=true]{hyperref}
\usepackage{xcolor}
\usepackage{amssymb}
\usepackage{stackengine,graphicx}
\usepackage{amsmath}
\usepackage{hyperref}
\usepackage{color}
\usepackage{stackengine}
\usepackage{subfigure}
\usepackage{verbatim}


\newcommand{\RNum}[1]{\uppercase\expandafter{\romannumeral #1\relax}}

\newcommand{\balancecolsandclearpage}{%
	\close@column@grid
	\cleardoublepage
	\twocolumngrid
}

\begin{document}

\title{Tailoring higher-order topological phases via orbital hybridization}

\author{Maxim Mazanov}
\email{maxim.mazanov@metalab.ifmo.ru}
\affiliation{School of Physics and Engineering, ITMO University, St. Petersburg, Russia}
\author{Maxim A. Gorlach}
\email{m.gorlach@metalab.ifmo.ru}
\affiliation{School of Physics and Engineering, ITMO University, St. Petersburg, Russia}

\begin{abstract}
    Higher-order topological insulators (HOTIs) have attracted much attention in photonics due to the tightly localized disorder-robust corner and hinge states. Here, we reveal an unconventional HOTI phase with vanishing dipole and quadrupole polarizations. This phase  arises in the array of evanescently coupled waveguides hosting degenerate $s$- and $d$-type orbital modes arranged in a square lattice with four waveguides in the unit cell. As we prove, the degeneracy of the modes with the different symmetry gives rise to the nontrivial topological properties rendering the system equivalent to the two copies of anisotropic two-dimensional Su-Schrieffer-Heeger model rotated by 90$^\circ$ with respect to each other and based on $s\pm d$ hybridized orbitals. Our results introduce a route to tailor higher-order band topology leveraging both crystalline symmetries and accidental degeneracies of the different orbital modes.

\end{abstract}

\date{\today}

\maketitle


\textit{Introduction}.--- Higher-order topological insulators (HOTIs) are unusual states of matter hosting topologically protected localized states of at least two dimensions less then the system itself~\cite{Schindler2018,Xie_HOTI_review_2021}. The understanding of their physics relies either on nontrivial bulk dipole polarization~\cite{Quantization_2019} or higher-order multipole polarization~\cite{Benalcazar_2017_Science, Benalcazar_2017_PRB} associated with the Bloch bands. Recent photonic realizations are exploiting suitably designed lattices~\cite{Wu_Hu,Noh2018}, intentionally tailored negative effective couplings or artificial gauge fields~\cite{Peterson2018,Mittal_2019,Hassan2019}, and the entire area continues to rapidly evolve~\cite{Xie_HOTI_review_2021}. In these approaches, lattice symmetry and the engineered couplings between the elementary building blocks play the key role ensuring the nontrivial topology of the bands.

In this Letter, we take another route to higher-order topology exploiting multi-mode constituent elements and harnessing {\it both} lattice geometry and accidental degeneracy of the different orbital modes. Previously, the concept of orbital hybridization has been used to tailor the dispersion inducing such features as Dirac cones and flat bands~\cite{Huang_2011, Sakoda_2012_3} or manipulate couplings between the elements by hybridizing either $s$ and $p$-type modes with the different inversion parities~\cite{SavelevGorlach_PRB, Aravena_PRA_2020, Li_2016, Li_2013, Pelegri_PRA_2019, Yin_superfluid_2015} or $p_x$ and $p_y$ modes corresponding to the different polarization states~\cite{Wu_PRL_2007,Poddubny2014,Bloch_PRL_2014, Bloch_PRL_2017, Bloch_PRX_2019}.

\begin{figure}[t!]
	\centering
	\includegraphics[width=0.4\textwidth]{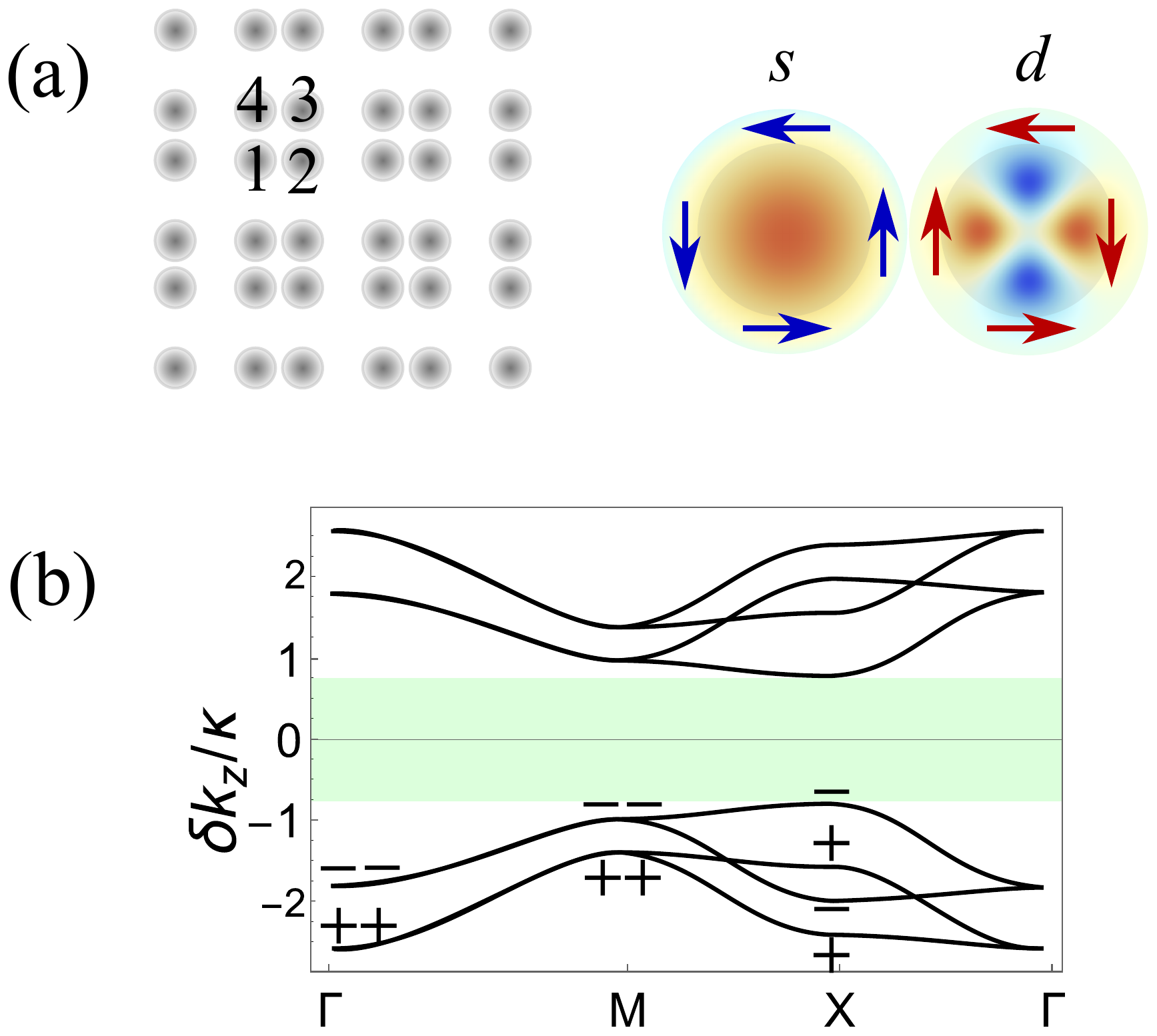}
	\caption{
	    (a) Two-dimensional Su-Schrieffer-Heeger lattice with degenerate monopolar (TE$_{0m}$, left) and quadrupolar (HE$_{2k}$, right) modes at each site. Arrows indicate the direction of electric field in the respective modes. 
		(b) Bulk band structure with a complete gap highlighted by green; $\pm$ signs indicate inversion eigenvalues at high-symmetry points. The calculation is performed for $\kappa=\gamma, \, \Delta=0.7\kappa$, $\alpha = 0.3$. 
	}
	\label{Fig1}
\end{figure}

In contrast, here we harness the idea of orbital hybridization to construct an unconventional HOTI phase arising in a square waveguide lattice with four sites in the unit cell [Fig.~\ref{Fig1}(a)] resembling the two-dimensional (2D)  Su-Schrieffer-Heeger model (SSH)~\cite{1DSSH,ZeroBerry_PRB, ZeroBerry_PRL}. While in the single orbital limit this model does not have zero energy bandgap, the degeneracy of monopolar ($s$-like) and quadrupolar ($d$-like) modes at each site gives rise to the complete bandgap hosting disorder-robust midgap corner states.

Unlike previous realizations of higher-order topology, this system is characterized by vanishing dipole and quadrupole polarizations. Despite that, we reveal the topological nature of the model by showing its equivalence to the direct sum of two anisotropic 2D SSH models for $s+d$ and $s-d$ hybrid orbitals. The effective anisotropy of the coupling for $s\pm d$ orbitals is induced via the generalized near-field Kerker effect~\cite{Savelev2019,Kivshar_OE_2018}. In turn, each of the hybrid orbitals treated as a distinct pseudospin has nontrivial topological properties giving rise to the topological corner states.

It should be also stressed that our system is different from the celebrated spinful $s-d$-hybridized model of quadrupole insulator reported in condensed matter physics~\cite{Strong_and_fragile_2020}. Contrary to that model relying on spin-orbit coupling, we introduce here geometric gap-opening terms which enable higher-order topology readily attainable in waveguide lattices.

\textit{Model and band structure.}~--- We consider a square lattice of two-mode waveguides depicted in Fig.~\ref{Fig1}(a), each of the waveguides hosts two modes having the same propagation constants in the chosen frequency range. $s$ and $d$ orbital modes correspond to TE$_{0m}$/TM$_{0n}$ and HE$_{2k}$ waveguide modes with arbitrary radial indices $m,n,k$, respectively, and typical field profiles sketched in Fig.~\ref{Fig1}(a).

The described model possesses $C_2$ rotational symmetry as well as inversion symmetry. Note that although the quadrupolar mode is doubly degenerate, its $45^\circ$-rotated partner does not contribute to the effective photonic Hamiltonian since its electric field has zero overlaps with the monopolar modes in the nearest-neighbour waveguides. However, in the case of triple degeneracy of TE$_{0m}$, TM$_{0n}$ and $2\times$HE$_{2k}$ orbitals we will have two noninteracting copies of our model.

Assuming the interaction of the nearest neighbors, we obtain the set of coupled-mode equations for the described two-mode lattice model (see Supplemental Materials \cite{Supplement} for the derivation): 
\begin{eqnarray}
\label{coupled_mode_eqns}
-i \frac{d}{dz} \begin{pmatrix} u \\ v \end{pmatrix}_{m,n}
&=&
\hat{\kappa}_- \left[ \begin{pmatrix} u \\ v \end{pmatrix}_{m-1,n}
+ \begin{pmatrix} u \\ v \end{pmatrix}_{m+1,n} \right]
+ 
\nonumber \\
\label{couplings}
&&\hat{\kappa}_+ \left[ \begin{pmatrix} u \\ v \end{pmatrix}_{m,n-1}
+ \begin{pmatrix} u \\ v \end{pmatrix}_{m,n+1} \right] 
,
\end{eqnarray}
where $z$ is the coordinate along the waveguide axes, $m$ and $n$ are horizontal and vertical unit cell indices, respectively, while $ u $ and $ v $ denote the amplitudes of the first (monopolar) and the second (quadrupolar) modes in each waveguide. Due to the coupling between the waveguides, these amplitudes depend on $z$.

Symmetries of the mode profiles result in the coupling matrices:
$ \hat{\kappa}_\pm = \left(
\begin{array}{cccccc}
\kappa & \pm i \Delta \\
\mp i \Delta & \gamma \\
\end{array}
\right) $, where $ \kappa $, $ \gamma $, $ \Delta $ are real-valued (positive) coupling integrals quantifying the interaction of the mode pairs $ {u,u} $~($ \kappa $), $ {u,v} $~($ \pm i\Delta $), and $ {v,v} $~($ \gamma $). 

Next, we incorporate geometric shrinking-expanding lattice distortions into our model [Fig.~\ref{Fig1}(a)]. For simplicity, we approximate the effect of such distortion by the single scalar parameter $0 \leq \alpha \leq 1$ that changes all the couplings proportionally, $\hat{\kappa}_\pm \rightarrow \alpha \hat{\kappa}_\pm $. Choosing the numbering of sites consistent with Fig.~\ref{Fig1}(a), we recover $8\times 8$ Bloch Hamiltonian
\begin{eqnarray}
    \label{Hamk1}
    \hat{H}(\textbf{k}) 
    = 
    \left(
    \begin{array}{cccc}
    0 & \beta_x^* \hat{\kappa}_- & 0 & \beta_y^* \hat{\kappa}_+ \\
    \beta_x \hat{\kappa}_- & 0 & \beta_y^* \hat{\kappa}_+ & 0 \\
    0 & \beta_y \hat{\kappa}_+ & 0 & \beta_x \hat{\kappa}_- \\
    \beta_y \hat{\kappa}_+ & 0 & \beta_x^* \hat{\kappa}_- & 0 \\
    \end{array}
    \right)
    ,
\end{eqnarray}
where $\beta_{x,y} \equiv \alpha e^{i k_{x,y}} +1$. Even though $\alpha$ should be generally regarded as a matrix with norm $||\hat{\alpha}|| < 1$, this simplified model already captures the topological origin of our system predicting topological corner states in agreement with the full-wave numerical simulations  discussed below.

The Bloch modes of the system are found as the solutions to the eigenvalue equation of the form
$
\label{bloch_eqn}
\hat{H}(\textbf{k})
(\textbf{v}_1, \textbf{v}_2, \textbf{v}_3, \textbf{v}_4)^T
=
\delta k_z
(\textbf{v}_1, \textbf{v}_2, \textbf{v}_3, \textbf{v}_4)^T
, 
$
where $\textbf{v}_i = (u_i, v_i)^T$, and the propagation constant plays the role of energy, as suggested by the analogy between the paraxial approximation of Maxwell's equations and the Shr{\"o}dinger equation~\cite{Lederer2008,Rechtsman2013}. 

The calculated band diagram for $\hat{H}(\textbf{k})$ with $\alpha<1$ is shown in Fig.~\ref{Fig1}(b). We observe that the 2D SSH-type distortions open a complete gap between the two sets of bands. In the special case $\Delta = \kappa = \gamma$, the bands become pairwise degenerate.

\textit{Higher-order topology.}~--- To assess the topological nature of our model, we first note that it possesses two commuting reflection symmetries $M_x$ and $M_y$~\cite{Supplement},
which rules out the higher-order quadrupolar topology since the bulk Wannier bands are then necessarily gapless~\cite{Benalcazar_2017_Science,Benalcazar_2017_PRB}.

Next, we test if our model is a higher-order topological crystalline insulator by calculating symmetry indicators at high-symmetry points of the Brillouin zone~\cite{Quantization_2019}. In this way, we find that the conventional HOTI index $\chi^{(2)}$ vanishes~\cite{Supplement}.

\begin{figure}[t!]
	\centering
	\includegraphics[width=0.48\textwidth]{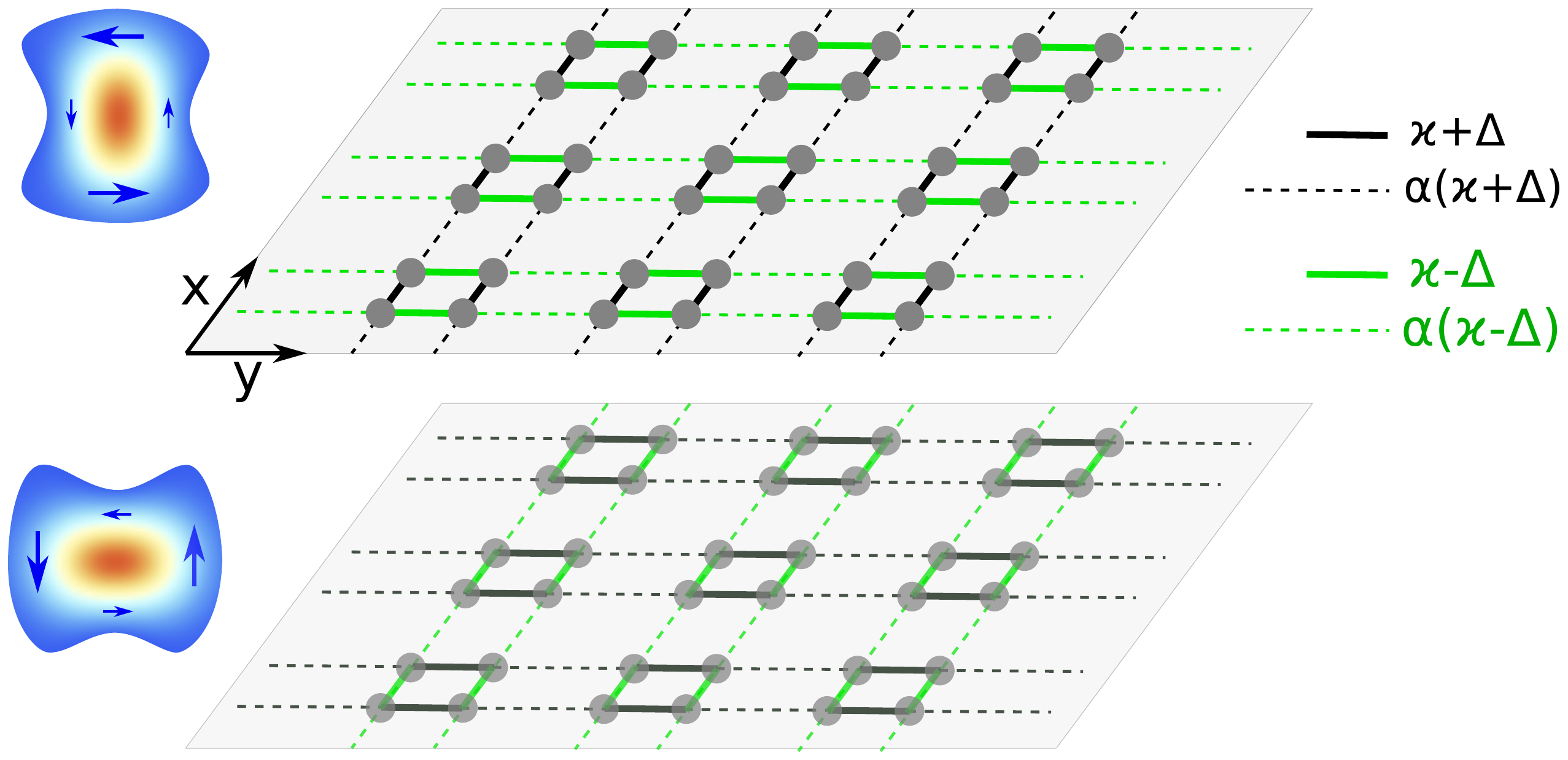}
	\caption{
		Equivalent two-layer model of our system. Each layer corresponds to the anisotropic 2D SSH model with the couplings indicated in the legend for the $s\pm d$ hybrid orbitals with the field profiles depicted schematically on the left of each layer. 
	}
	\label{Fig2}
\end{figure}

Despite that, our system still represents the topological phase being a sum of two known topologically nontrivial primitive generators~\cite{Quantization_2019}. In particular, in the special case $\kappa=\gamma$, the Hamiltonian Eq.~\eqref{Hamk1} can be transformed to the following block-diagonal form: 
\begin{eqnarray}
\label{H'main}
    \hat{H}' 
    \equiv 
    \hat{U}^\dag \hat{H} \hat{U} 
    = 
    \left(
    \begin{array}{cc}
    \hat{h}(k_x,k_y) & 0 \\
    0 & \hat{h}(k_y,-k_x) \\
    \end{array}
\right) 
\end{eqnarray}
with the basis formed by $s+d$ orbitals for the upper diagonal block, and by $s-d$ orbitals for the lower block~\cite{Supplement}. 
The matrix block $\hat{h}(k_x,k_y)$ in Eq.~\eqref{H'main} represents the anisotropic 2D SSH model with real couplings $\kappa+\Delta$, $\alpha(\kappa+\Delta)$ along $x$, and $\kappa-\Delta$, $\alpha(\kappa-\Delta)$ along $y$, see Fig.~\ref{Fig2}. 
When $\alpha < 1$ and $\Delta \neq \kappa$, such anisotropic 2D SSH model 
is a well-known $C_2$-symmetric higher-order topological crystalline insulating phase~\cite{Benalcazar_2017_PRB,Xie_2018}, corresponding to a primitive generator $h^{(4)}_{1b}$~\cite{Quantization_2019} with topological index $\chi^{(2)}= (-1,-1,0)$, bulk polarization $P = (1/2; 1/2)$, zero bulk quadrupole moment and alternating corner charges $\pm e/2$ at half-filling~\cite{Benalcazar_2017_Science}. These corner charges are quantized by the two commuting reflection symmetries $M_x$ and $M_y$~\cite{Benalcazar_2017_PRB,Xie_2018}.

\begin{figure}[t!]
	\centering
	\includegraphics[width=0.48\textwidth]{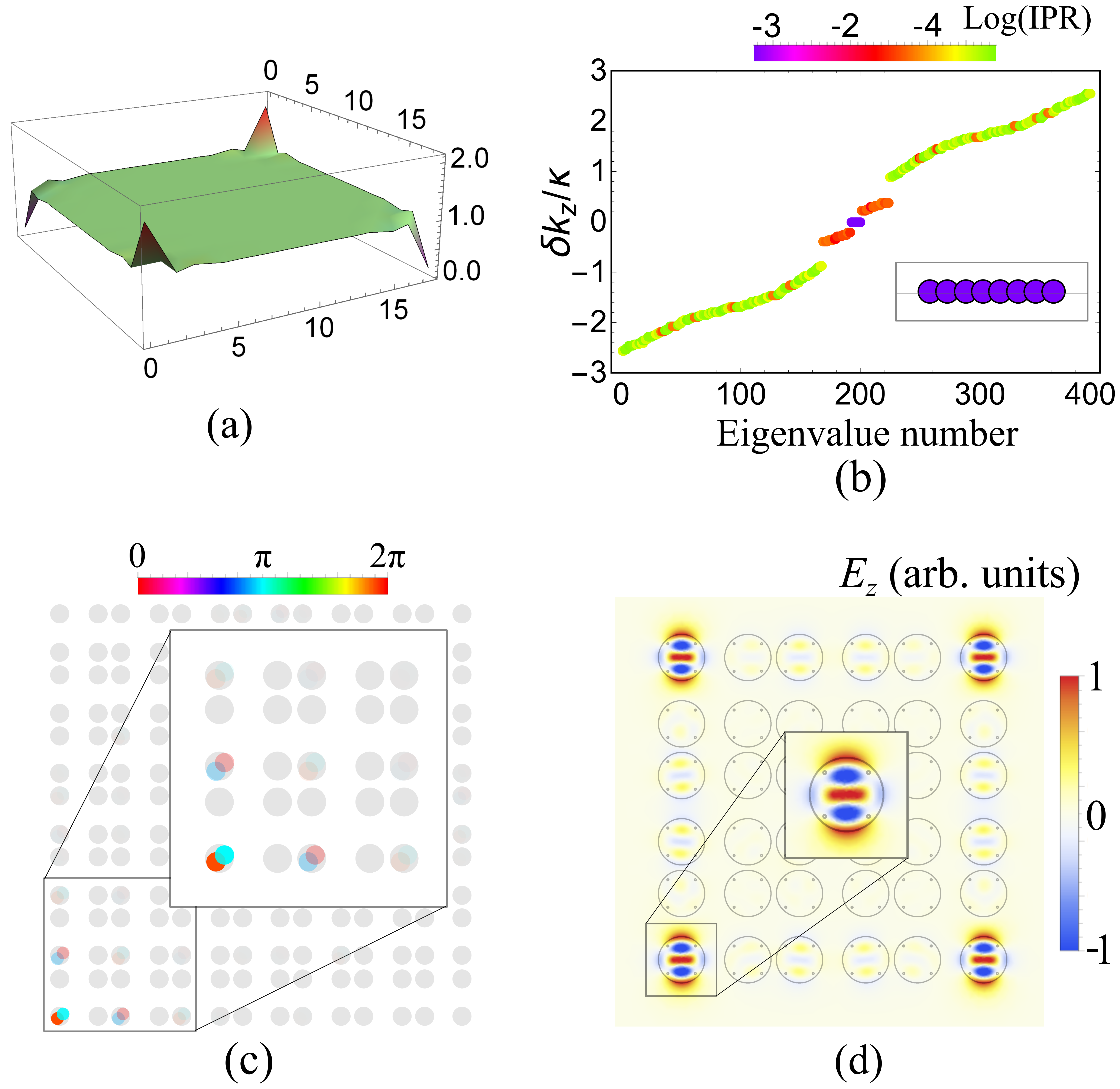}
	\caption{
	    (a) Charge distribution at half-filling calculated for the finite lattice $14\times 14$ sites with the same parameters as in Fig.~\ref{Fig1}. Corner charges $Q_{{\rm corner}} = \pm 1$ are obtained by the summation over several sites nearest to the corner. In this calculation, on-site energies of the diagonal sites at each unit cell were set to small $\delta = \pm 10^{-2} \kappa$ as proposed in Ref.~\cite{Benalcazar_2017_PRB}. 
		(b) Spectrum of the same finite  lattice. The logarithm of inverse participation ratio is color-coded. The inset shows $8$-fold degenerate midgap corner states. 
		(c) Real-space profile for one of the corner modes. Two circles in each waveguide indicate the amplitudes of the two modes, and the phase is color-coded. 
		(d) The map of longitudinal component $E_z$ of electric field for the corner states with propagation constants $245 \,\text{rad/m}$ obtained from the full-wave numerical simulations. Each waveguide~\cite{Supplement} hosts TM$_{02}$ and HE$_{22}$ degenerate modes, has radius $1.26 \,\text{cm}$ and permittivity $\varepsilon = 9.9$ corresponding to commercially available ceramic aluminium oxide. The center-to-center nearest-neighbour distance in expanded unit cell is $3.6\,\text{cm}$, and periods in $x$ and $y$ directions are $6.44\,\text{cm}$. The calculations are performed for a fixed frequency $f = 8.83 \,\text{GHz}$ using COMSOL Multiphysics software package.
	}
	\label{Fig3}
\end{figure}

Thus, in the limiting case $\kappa=\gamma$ our model represents a pair of two  noninteracting anisotropic 2D SSH layers, one rotated by $90^\circ$ with respect to another (Fig.~\ref{Fig2}), which corresponds to the direct sum $h^{(4)}_{1b}\bigoplus h^{(4)}_{1b}$ of the primitive generators. 
The index theorem~\cite{Quantization_2019} then implies that the topological index of our model is the sum of indices of its primitive generators, $\chi^{(2)}= (-1,-1,0)+(-1,-1,0) = (-2,-2,0)$, and the corner charge is also summed, $Q_{{\rm corner}} = 2\cdot e/2 = e$. Because of the pseudospin degree of freedom, integer corner charge is observable and gives rise to the pair of degenerate corner states with the distinct pseudospins, i.e. different phase shifts between $s$ and $d$ orbitals. The value of the corner charge is also confirmed by the direct calculation of the charge distribution at half-filling where the states with $\delta k_z \leq 0$ are considered [Fig.~\ref{Fig3}(a)]. When $\kappa \neq \gamma$, the two anisotropic 2D SSH layers couple~\cite{Supplement}. However, as long as the inter-layer coupling does not close the bulk gap, the topology of the whole system remains the same as in the case of non-interacting layers.

The mere fact that our model, while having trivial conventional HOTI index, is a sum of two nontrivial primitive generators for opposite pseudospins, allows for its further interpretation as a higher-order spin Hall insulator. Such interpretation also aligns with the physics of quantum spin Hall effect~\cite{Ozawa_RMP_2019} characterized by vanishing Chern number, but nonzero and opposite Chern numbers for the two decoupled spin subsystems.

\textit{Corner and edge states.}--- The higher-order topology of the system manifests itself when both constituent anisotropic 2D SSH layers are in their HOTI phases, Fig.~\ref{Fig3}(b,c). 
The topological corner states appear in the middle of the bulk gap with a typical field distribution shown in Fig.~\ref{Fig3}(c). 
To distinguish the modes by their localization, we calculate the two-mode generalization of the inverse participation ratio (IPR)~\cite{Thouless_1974}
\begin{equation}
\mathfrak{I}
=
\frac{
\sum_{m, n} \left( \left|u_{m n}\right|^{2} + \left|v_{m n}\right|^{2} \right)^2
}
{
\sum_{m, n} \left( \left|u_{m n}\right|^{2} + \left|v_{m n}\right|^{2} \right)
}
. 
\end{equation}
For different types of states, IPR shows different scaling with the size of the system $N$: $\mathfrak{I} \propto 1/N$ for the bulk states, $\mathfrak{I} \propto 1/\sqrt{N}$ for edge states, and $ \mathfrak{I} \propto 1$ for corner states~\cite{Olekhno_2021,Kachin_2021}. Indeed, we recognize three different types of scaling inspecting $\ln (\mathfrak{I})$, which is color coded in panel~(b) of Fig.~\ref{Fig3}.

Furthermore, the corner states are pinned to zero energy by the generalized chiral (sublattice) symmetry~\cite{Weiner_generalized_2019} which ensures that the spectrum is symmetric~\cite{Supplement}. The localization length of the corner states measured in number of sites turns out to be $\lambda_{{\rm loc}} = 1 / |\ln \alpha|$, as in the one-dimensional SSH model~\cite{Supplement,1DSSH}. 
Note that these corner states are not the simultaneous end states of the SSH columns and rows (uncoupled at $\kappa = \gamma = \Delta$) since they host orthogonal orbital states.

To assess the robustness of predicted states, we also consider the effect of imperfect mode degeneracy and disorder in the couplings on the corner states. We find that the topology remains intact until the critical mode detuning which closes the bulk gap: $\delta k_{c} = 2 (1-\alpha ) \cdot \min( \gamma, \kappa )$, which is of the same order of magnitude as the couplings themselves ~\cite{Supplement}. Similarly, zero-energy corner states survive the general Hermitian disorder and retain their near-zero energy until they mix with the bulk states upon critical value of disorder comparable to the couplings~\cite{Supplement}.

In the limit $\Delta=\kappa$, the edge states shown in orange in Fig.~\ref{Fig3}(b) can be viewed as combinations of $2N$ edge states of uncoupled columns/rows with up/down orbital pseudospins, respectively, and thus inherit their topological properties directly from the one-dimensional SSH model. In particular, they are described by the same topological $\mathbb{Z}_2$ invariant (or Zak's phase) and appear at the edges connected to the bulk via the weak link. 
We note that while the SSH-like edge states in our model are different from the dispersive edge states in single-mode 2D SSH models~\cite{ZeroBerry_PRB, ZeroBerry_PRL, Xie_2018, DirectObs_PRL, PRApplied}, they are also protected by the non-trivial $\mathbb{Z}_2$ invariant along specific directions of the lattice, and have zero Chern number~\cite{Marques_2018}.

To confirm the predicted physics, we perform full-wave numerical simulations in microwaves, starting from a single waveguide with degenerate TM$_{02}$ and HE$_{22}$ modes. Despite the presence of the rotated HE$_{22}$ modes which do not interact with the modes of interest and the spectrally close orbital $f$-modes, our proposed $s$-$d$ model succeeds in capturing the topology of the system, and we indeed observe midgap corner states with the same propagation constant as that of the two degenerate orbitals. The calculated field profile of the corner mode is depicted in Fig.~\ref{Fig3}(d). The map of the longitudinal component $E_z$ of electric field, which is responsible for the inter-orbital coupling of the chosen modes, clearly shows $s + d$ hybridized modes tightly localized at the corners. Another four $s - d$ hybrid corner states not shown here are just $90^\circ$-rotated copies of the modes shown in Fig.~\ref{Fig3}(d). Further details regarding full-wave numerical simulations of our model are provided in Supplemental Materials~\cite{Supplement}.

\textit{Discussion and conclusions.}~--- In summary, we have demonstrated, both analytically and numerically, that higher-order topological states can be tailored in waveguide lattices by utilizing the interference of accidentally degenerate on-site orbital modes combined with the suitable lattice symmetry. As we prove, the combination of these mechanisms gives an access to the unconventional types of topology when the net bulk dipole and quadrupole polarizations vanish while having nonzero value for the fixed pseudospin.

Our calculations show that the proposed higher-order spin-Hall system can be readily realized for the arrays of evanescently coupled microwave waveguides. We anticipate, however, that the discussed physics can be also brought to the optical range by utilizing the arrays of laser-written waveguides which have recently been shown to exhibit sizable inter-orbital coupling~\cite{Silva_PRL_2021}. We believe that owing to the versatility and high tunability of photonic platforms, many more models and higher-order topological phases are yet to be discovered on this route. 

In a broader perspective, the discussed physics is also applicable to bipartite cold atom lattices~\cite{Li_2016}, where orbital hybridization is achieved by tuning the effective optical potential profile. We envision that our ideas of mode interference can be applied to engineer higher-order topology in three spatial dimensions benefitting from the multipole classification and symmetry analysis of the eigenmodes in electromagnetic resonators~\cite{Gladyshev_2020} combined with the vast knowledge of generalized Kerker effects~\cite{Kivshar_OE_2018}.

\textit{Acknowledgments.}~--- We acknowledge Roman Savelev and Daria Smirnova for valuable discussions. This work was supported by the Russian Science Foundation (grant No.~20-72-10065).


\bibliography{refs}

\end{document}


\title{Supplemental Materials: \\ Tailoring higher-order topological phases via orbital hybridization}

\author{Maxim Mazanov}
\email{mazanovmax@gmail.com}
\affiliation{School of Physics and Engineering, ITMO University, St. Petersburg, Russia}
\author{Maxim A. Gorlach}
\email{m.gorlach@metalab.ifmo.ru}
\affiliation{School of Physics and Engineering, ITMO University, St. Petersburg, Russia}

\maketitle

\widetext

\setcounter{equation}{0}
\setcounter{figure}{0}
\setcounter{table}{0}
\setcounter{page}{1}
\setcounter{section}{0}
\makeatletter
\renewcommand{\theequation}{S\arabic{equation}}
\renewcommand{\thefigure}{S\arabic{figure}}
\renewcommand{\bibnumfmt}[1]{[S#1]}
\renewcommand{\citenumfont}[1]{S#1}

\section{Supplemental Note 1: Details on coupled-mode equations}

Coupled-mode equations for an array of two-mode waveguides have been derived in Ref.~\cite{SavelevGorlach_PRB} (see Supplementary materials, Section~I of that paper). In that work, by using orthogonality of the modes of a single waveguide and the fact that the both modes exponentially decay outside of the waveguide, the following generalization of coupled-mode equations has been obtained from Maxwell's equations (here, we further generalize it to two-dimensional lattice): 
\begin{equation}
\frac{d}{d z}\left(\begin{array}{c}
a_{m,n}^{(1)} \\
a_{m,n}^{(2)}
\end{array}\right)=i \sum_{k,l=\pm 1}\left(\begin{array}{cc}
\kappa_{m,n; m+k, n+l}^{(11)} & \kappa_{m,n; m+k, n+l}^{(12)} \\
\kappa_{m,n; m+k, n+l}^{(21)} & \kappa_{m,n; m+k, n+l}^{(22)}
\end{array}\right)\left(\begin{array}{c}
a_{m+k,n+l}^{(1)} \\
a_{m+k,n+l}^{(2)}
\end{array}\right)
, 
\end{equation}
where $a_{m, n}^{(1,2)}(z)$ are the amplitudes of the two modes in coupled waveguides, and the elements of the coupling matrices are defined as coupling integrals over the volume of the unit cell: 
\begin{equation}
\kappa_{m,n; m+k, n+l}^{(p q)}=\frac{\omega}{c} \frac{1}{\mathcal{P}^{(p)}_{m,n}} \int\left[\varepsilon(\mathbf{r})-\varepsilon_{m+k,n+l}(\mathbf{r})\right] \mathbf{E}_{m,n}^{(p) *} \cdot \mathbf{E}_{m+k,n+l}^{(q)} d V
,
\end{equation}
where $\omega$ is the frequency of interest, $ \epsilon (r) $ is the permittivity distribution in the entire array consisting of waveguides, $ \epsilon_{m+k,n+l} (r) $ is the permittivity distribution in the waveguide, while $\mathcal{P}_{m,n}^{(p)}=\hat{\mathbf{z}} \cdot \int\left[\mathbf{E}_{m,n}^{(p) *} \times \mathbf{H}_{m,n}^{(p)}+\mathbf{E}_{m,n}^{(p)} \times \mathbf{H}_{m,n}^{(p) *}\right] d V$ is proportional to the power carried by the respective waveguide mode and can be set to unity by the proper normalization of the mode fields; $p$, $q$ indices enumerate the modes of a waveguide being equal to 1 or 2; $k,l = \pm 1$.

Although the magnitude of the off-diagonal terms in the coupling matrices $\kappa_{m,n; m+k, n+l}^{(p q)}$ is affected by the relative normalization of the modes, by choosing a specific relative normalization we can ensure that the coupling matrix for each pair of neighboring waveguides is symmetric/antisymmetric~\cite{SavelevGorlach_PRB}.

The structure of the coupling matrices could be deduced from the mode symmetry using the schematic mode profiles shown in Fig.~1(a) in the main text. From the definition of the coupling matrices, it follows that $\kappa_{m,n; m+k,n+l}^{(p q)} \propto \textbf{E}_{m,n}^{(p) *} \textbf{E}_{m+k,n+l}^{(q)}$. Considering only nearest-neighbour coupling, we recognize that all couplings are governed by two Hermitian coupling matrices: 
\begin{eqnarray}
\hat{\kappa}_\pm = \left(
\begin{array}{cccccc}
\kappa & \pm i \Delta \\
\mp i \Delta & \gamma \\
\end{array}
\right)
, 
\end{eqnarray}
with $\kappa_{m,n; m+1,n}^{(p q)} = \kappa_{m,n; m-1,n}^{(p q)} = \hat{\kappa}_{-} $, and $\kappa_{m,n; m,n+1}^{(p q)} = \kappa_{m,n; m,n-1}^{(p q)} = \hat{\kappa}_{+} $. These matrices have been used in Eq.~(1) and further in the main text.

\section{Supplemental Note 2: Generalized chiral and mirror symmetries}

The Bloch Hamiltonian $\hat{H}(\textbf{k})$ [Eq.~(2) in the main text] possesses generalized chiral symmetry~\cite{Weiner_generalized_2019} for any choice of the couplings. The proof is given simply by constructing a respective unitary generalized chiral symmetry operator, 
\begin{eqnarray}
\hat{\Gamma}
=
\begin{pmatrix}
	\hat{I} & 0 & 0 & 0 \\
	0 & e^{i \pi/4} \cdot \hat{I} & 0 & 0 \\
	0 & 0 & e^{i 2\pi/4} \cdot \hat{I} & 0 \\
	0 & 0 & 0 & e^{i 3\pi/4} \cdot \hat{I} \\
\end{pmatrix}
,
\end{eqnarray}
where $\hat{I}$ is a $2\times 2$ unit matrix, and zeros stand for the $2\times 2$ zero matrices (note that $\hat{\Gamma}^4 = \hat{I}$). Then, it is straightforward to check that 
\begin{eqnarray}
\hat{H}(\textbf{k})
+
\hat{\Gamma}
\hat{H}(\textbf{k})
\hat{\Gamma}^{-1} 
+
\hat{\Gamma}^2 
\hat{H}(\textbf{k})
\hat{\Gamma}^{-2} 
+
\hat{\Gamma}^3 
\hat{H}(\textbf{k})
\hat{\Gamma}^{-3} 
=
0
, \nonumber
\end{eqnarray}
which proves that $\hat{H}(\textbf{k})$ possesses generalized chiral symmetry. This explains why the calculated spectrum is symmetric. 
The topological corner states described in the main text are pinned to zero energy exactly by this chiral symmetry as long as next-nearest neighbor couplings are negligible. 
Moreover, all corner states may be obtained by successively applying the generalized chiral symmetry operator $\hat{\Gamma}$ to one known corner state.

Our model also possesses two mirror symmetries, which in the same basis read: 

\begin{eqnarray}
\hat{M}_x
=
\pm \begin{pmatrix}
	0 & \hat{I} & 0 & 0 \\
	\hat{I} & 0 & 0 & 0 \\
	0 & 0 & 0 & \hat{I} \\
	0 & 0 & \hat{I} & 0 \\
\end{pmatrix}
, \quad
\hat{M}_y
=
\pm \begin{pmatrix}
	0 & 0 & 0 & \hat{I} \\
	0 & 0 & \hat{I} & 0 \\
	0 & \hat{I} & 0 & 0 \\
	\hat{I} & 0 & 0 & 0 \\
\end{pmatrix}
\end{eqnarray}
where $\hat{I}$ is the $2 \times 2$ identity matrix, and $+$ or $-$ stands for the $\{$TM,HE$\}$ or $\{$TE,HE$\}$ pair of degenerate interacting modes, respectively. 
Since these symmetry operators commute, $[\hat{M}_x, \hat{M}_y] = 0$, for both choices of two interacting orbital modes, the  Wannier bands are gapless and the quadrupole moment thus necessarily vanishes, as discussed in the main text.

\section{Supplemental Note 3: Equivalence to a pair of uncoupled anisotropic SSH models for $\kappa=\gamma$}

The Hamiltonian $\hat{H}(\textbf{k})$ [Eq.~(2) in the main text] at $\kappa=\gamma$ takes the block-diagonal form \begin{eqnarray}
\label{H'}
    \hat{H}' 
    \equiv 
    \hat{U}^\dag_1 \hat{H} \hat{U}_1 
    = 
    \left(
    \begin{array}{cc}
    \hat{h}(k_x,k_y) & 0 \\
    0 & \hat{h}(k_y,-k_x) \\
    \end{array}
\right) 
\end{eqnarray}
with the help of the following unitary transformation: 
\begin{eqnarray}
\hat{U}_1
=
\frac{1}{2}
\left(
\begin{array}{cccccccc}
 0 & 0 & 0 & 1+i & 0 & 1-i & 0 & 0 \\
 0 & 0 & 0 & 1-i & 0 & 1+i & 0 & 0 \\
 0 & 0 & 1+i & 0 & 0 & 0 & 0 & 1-i \\
 0 & 0 & 1-i & 0 & 0 & 0 & 0 & 1+i \\
 1+i & 0 & 0 & 0 & 0 & 0 & 1-i & 0 \\
 1-i & 0 & 0 & 0 & 0 & 0 & 1+i & 0 \\
 0 & 1+i & 0 & 0 & 1-i & 0 & 0 & 0 \\
 0 & 1-i & 0 & 0 & 1+i & 0 & 0 & 0 \\
\end{array}
\right)
. 
\end{eqnarray}
In Eq.~\eqref{H'}, the matrix $\hat{h}(k_x,k_y)$ reads 
\begin{eqnarray}
\label{h_k}
    \hat{h}(k_x,k_y)
    = 
    \left(
\begin{array}{cccc}
 0 & \left(e^{i k_x} \alpha +1\right) (\kappa + \Delta) & \left(e^{i k_y} \alpha +1\right) (\kappa - \Delta) & 0 \\
 (e^{-i k_x} \alpha + 1) (\kappa + \Delta) & 0 & 0 & \left(e^{i k_y} \alpha +1\right) (\kappa - \Delta) \\
 \left(e^{-i k_y} \alpha +1\right) (\kappa - \Delta) & 0 & 0 & \left(e^{i k_x} \alpha +1\right) (\kappa + \Delta) \\
 0 & \left(e^{-i k_y} \alpha +1\right) (\kappa - \Delta) & (e^{-i k_x} \alpha + 1) (\kappa + \Delta) & 0 \\
\end{array}
\right)
. 
\end{eqnarray}
When $\Delta=\kappa$, each block~\eqref{h_k} separates into two non-interacting blocks, each representing a 1D SSH model with the alternating couplings $\kappa+\Delta$ and $\alpha(\kappa+\Delta)$, in agreement with the fact that at $\kappa=\gamma=\Delta$ the model separates into non-interacting 1D SSH columns, for RCP orbital pseudospin states, and rows, for LCP orbital pseudospin states. Indeed, the block $\hat{h}(k_x,k_y)$ represents the rows, and block $\hat{h}(k_y,-k_x)$ represents the columns. 

\begin{figure}[t!]
	\centering
	\includegraphics[width=0.98\textwidth]{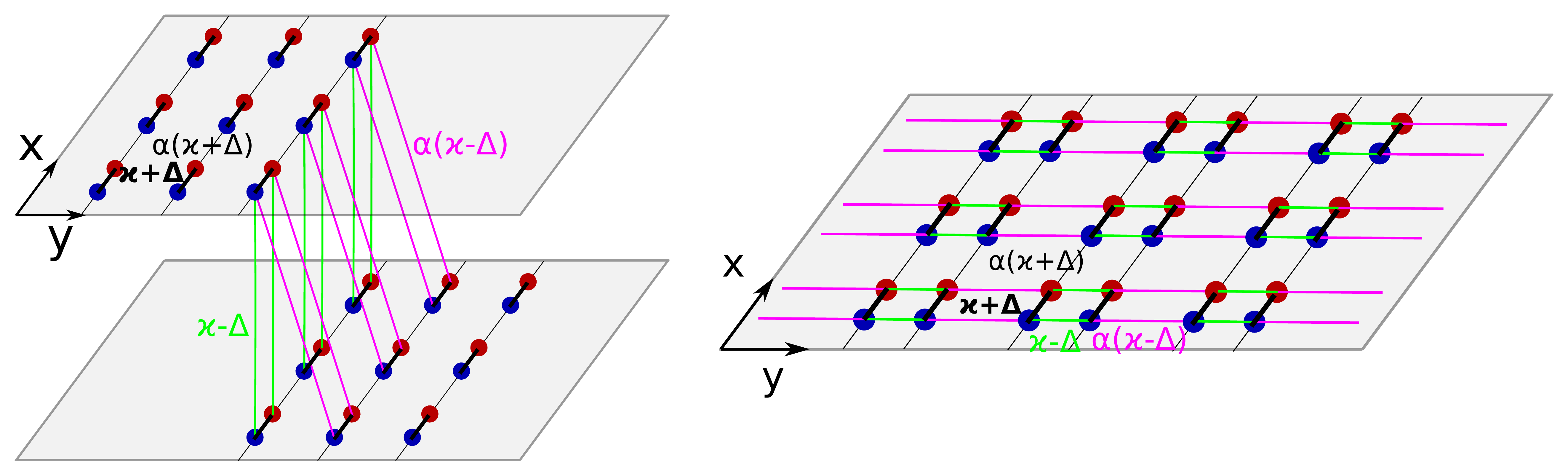}
	\caption{
		Two equivalent representations for the one of the two anisotropic 2D SSH layers of the equivalent model, represented by the block $\hat{h}(k_x,k_y)$ [Eq.~\eqref{h_k}]. 
	}
	\label{2DanizSSH}
\end{figure}

When $\Delta \neq \kappa$, the rows and the columns couple separately within each block, but the blocks $\hat{h}(k_x,k_y)$ and $\hat{h}(k_y,-k_x)$ remain non-interacting (i.e., there is still no interaction of LCP and RCP states). 
Looking closer at the structure of additional couplings in the blocks~\eqref{h_k}, we recognize that they represent the anisotropic 2D SSH model with couplings $\kappa+\Delta$ and $\alpha(\kappa+\Delta)$ along $x$, and $\kappa-\Delta$ and $\alpha(\kappa-\Delta)$ along $y$, see Fig.~\ref{2DanizSSH}. The anisotropic 2D SSH model with such configuration of couplings 
is a well-known $C_2$-symmetric higher-order topological crystalline insulator  phase~\cite{Xie_2018}, corresponding to a primitive generator $h^{(4)}_{1b}$~\cite{Quantization_2019} with bulk polarization $P = (1/2; 1/2)$ and alternating corner charges $\pm e/2$ at half-filling when an infinitely small staggered onsite potential slightly breaking the inversion symmetry is induced~\cite{Benalcazar_2017_Science}. 
These corner charges are quantized by two commuting reflection symmetries $M_x$ and $M_y$~\cite{Xie_2018}. 
Our model in the limiting case $\kappa=\gamma$ [Eq.~\eqref{H'}] thus represents two such noninteracting anisotropic 2D SSH ``layers'', one rotated at $90^\circ$ with respect to another, 
which corresponds to the direct sum $h^{(4)}_{1b}\bigoplus h^{(4)}_{1b}$ of the primitive generators. 

When $\kappa \neq \gamma$, the two anisotropic 2D SSH layers (with RCP and LCP pseudospin states) couple. This coupling could be thus viewed as a particular form of pseudospin-orbit interaction. 
However, as long as this inter-layer coupling does not close the bulk gap, the topology of the whole system remains the same as in the case of non-interacting layers.

\section{Supplemental Note 4: Symmetry indicators and the topological index}

As discussed in Ref.~\cite{Quantization_2019}, the bulk polarization $\textbf{P}$ for $C_2$-symmetric models is given by
\begin{equation}
    \mathbf{P}^{(2)}=\frac{e}{2}([Y_{1}^{(2)}]+[M_{1}^{(2)}]) \mathbf{a}_{1}+\frac{e}{2}([X_{1}^{(2)}]+[M_{1}^{(2)}]) \mathbf{a}_{\mathbf{2}} \mspace{8mu} \text{mod}\,e
    , 
\end{equation}
where $\mathbf{a}_{\mathbf{1,2}}$ are lattice translation vectors, and $[\Pi_{1}^{(2)}] \equiv \# \Pi_{1}^{(2)}-\# \Gamma_{1}^{(2)}$ are the integer topological invariants where $\# \Pi_{1}^{(2)}$ is the number of energy bands odd under $C_2$ rotation below zero energy at high-symmetry point $\Pi \in \{\Gamma,X,Y,M\}$ in the Brillouin zone.

The $C_2$ rotation operator in the basis of sites shown in Fig.~1(a) in the main text reads
\begin{eqnarray}
    \hat{R}_2
    =
    \left(
    \begin{array}{cccccccc}
    0 & 0 & 0 & 0 & 1 & 0 & 0 & 0 \\
    0 & 0 & 0 & 0 & 0 & 1 & 0 & 0 \\
    0 & 0 & 0 & 0 & 0 & 0 & 1 & 0 \\
    0 & 0 & 0 & 0 & 0 & 0 & 0 & 1 \\
    1 & 0 & 0 & 0 & 0 & 0 & 0 & 0 \\
    0 & 1 & 0 & 0 & 0 & 0 & 0 & 0 \\
    0 & 0 & 1 & 0 & 0 & 0 & 0 & 0 \\
    0 & 0 & 0 & 1 & 0 & 0 & 0 & 0 \\
\end{array}
\right)
.
\end{eqnarray}
It is straightforward to check that $\hat{R}_2$ commutes with the Hamiltonian Eq.~(2) of the main text at high-symmetry points $\Gamma$, $X$, $Y$ and $M$. 

Next, we calculate the $\hat{R}_2$ eigenvalues for the eigenstates below zero energy. For the calculation, we may take any coupling parameters as long as the bulk bandgap is open. 
First, we pick the cell with weak links, $\gamma=\kappa=1, \, \Delta=0.7, \, \alpha = 0.3$ that correspond to Fig.~1(b) in the main text. 
The obtained energy-eigenvalue pairs are summarized in the following table (note that the $\hat{R}_2$ rotation eigenvalues at $Y$ point turn out to be the same as for the $X$ point):  
\begin{table}[H]\centering
\begin{tabular}{lll}
$\varepsilon$ & $p\left(R_{2}\right)$ \\
\toprule
$\Gamma \text{-point } \quad\left(k_{x}=0, k_{y}=0\right):$ \\
$\varepsilon_{1}=-2.6$ & $1$ \\
$\varepsilon_{2}=-2.6$ & $1$ \\
$\varepsilon_{3}=-1.82$ & $-1$ \\
$\varepsilon_{4}=-1.82$ & $-1$ \\
\midrule
$X \text{-point } \quad\left(k_{x}=\pi, k_{y}=0\right):$ \\
$\varepsilon_{1}=-2.42$ & $1$ \\
$\varepsilon_{2}=-2.0$ & $-1$ \\
$\varepsilon_{3}=-1.58$ & $1$ \\
$\varepsilon_{4}=-0.8$ & $-1$ \\
\midrule
$Y \text{-point } \quad\left(k_{x}=0, k_{y}=\pi \right):$ \\
$\varepsilon_{1}=-2.42$ & $1$ \\
$\varepsilon_{2}=-2.0$ & $-1$ \\
$\varepsilon_{3}=-1.58$ & $1$ \\
$\varepsilon_{4}=-0.8$ & $-1$ \\
\midrule
$M \text{-point } \quad\left(k_{x}=\pi, k_{y}=\pi \right):$ \\
$\varepsilon_{1}=-1.4$ & $1$ \\
$\varepsilon_{2}=-1.4$ & $1$ \\
$\varepsilon_{3}=-0.98$ & $-1$ \\
$\varepsilon_{4}=-0.98$ & $-1$ \\
\bottomrule
\end{tabular}
\end{table}

Next, we pick the cell with strong links, $\gamma=\kappa=1, \, \Delta=0.7, \, \alpha = 1.7$. We find: 
\begin{table}[H]\centering
\begin{tabular}{lll}
$\varepsilon$ & $p\left(R_{2}\right)$ \\
\toprule
$\Gamma \text{-point } \quad\left(k_{x}=0, k_{y}=0\right):$ \\
$\varepsilon_{1}=-5.4$ & $1$ \\
$\varepsilon_{2}=-5.4$ & $1$ \\
$\varepsilon_{3}=-3.78$ & $-1$ \\
$\varepsilon_{4}=-3.78$ & $-1$ \\
\midrule
$X \text{-point } \quad\left(k_{x}=\pi, k_{y}=0\right):$ \\
$\varepsilon_{1}=-4.8$ & $-1$ \\
$\varepsilon_{2}=-4.38$ & $1$ \\
$\varepsilon_{3}=-2.0$ & $-1$ \\
$\varepsilon_{4}=-0.38$ & $1$ \\
\midrule
$Y \text{-point } \quad\left(k_{x}=0, k_{y}=\pi \right):$ \\
$\varepsilon_{1}=-4.8$ & $-1$ \\
$\varepsilon_{2}=-4.38$ & $1$ \\
$\varepsilon_{3}=-2.0$ & $-1$ \\
$\varepsilon_{4}=-0.38$ & $1$ \\
\midrule
$M \text{-point } \quad\left(k_{x}=\pi, k_{y}=\pi \right):$ \\
$\varepsilon_{1}=-1.4$ & $1$ \\
$\varepsilon_{2}=-1.4$ & $1$ \\
$\varepsilon_{3}=-0.98$ & $-1$ \\
$\varepsilon_{4}=-0.98$ & $-1$ \\
\bottomrule
\end{tabular}
\end{table}

\noindent Thus, in both cases we find $[X_{1}^{(2)}] = [Y_{1}^{(2)}] = [M_{1}^{(2)}] = 0$, hence the bulk polarization vanishes, $\textbf{P} = (0,0)$, and the topological index for $C_2$-symmetric crystalline insulators~\cite{Quantization_2019} is trivial, $\chi^{(2)}=([X_{1}^{(2)}],[Y_{1}^{(2)}],[M_{1}^{(2)}]) = (0,0,0)$.

However, as our model corresponds to the direct sum of two primitive generators for each pseudospin, $h^{(4)}_{1b}\bigoplus h^{(4)}_{1b}$, each having a topological index $\chi^{(2)}= (-1,-1,0)$ and corner charges $Q_{corner} = e/2$~\cite{Quantization_2019}, it is rather this topological index which defines the topology of the whole model than the trivial index calculated for all bands. 
More explicitly, the index theorem for Hamiltonians being the direct sum of several primitive generators, 
defined in Ref.~\cite{Quantization_2019}, implies that for our model, the topological index is the sum of those for primitive generators, $\chi^{(2)}= (-1,-1,0)+(-1,-1,0) = (-2,-2,0)$, and the corner charge is also summed, $Q_{{\rm corner}} = 2\cdot e/2 = e$, which is confirmed in our direct calculation presented in panel (d) of Fig.~1 in the main text.

\section{Supplemental Note 5: Localization length of the corner state}

Here we calculate the localization length of the corner state from the tight-binding equations. 
In the basis of sites shown in Fig.~1(b) in the main text, we try the anzatz exponentially localized in lower left lattice corner with only the $3^{\text{rd}}$ site having nonzero amplitudes over the lattice: 
\begin{eqnarray}
    (u,v)^T_{m,n,3}
    &=&
    (u,v)^T_{0,0,3} \, (-1)^{m+n} \, e^{-(m+n)/\lambda_{{\rm loc}}}
    , \nonumber \\
    (u,v)^T_{m,n,1}
    &=&
    (u,v)^T_{m,n,2}=(u,v)^T_{m,n,4} = 0
    , 
\end{eqnarray}
where $\lambda_{{\rm loc}}$ is the localization length measured in number of sites. For simplicity, we consider the lattice extending infinitely from the corner, occupying the whole first quadrant in $\{m,n\}$ lattice coordinates. Considering the $4^{\text{th}}$ site in lattice cell $\{1,0\}$, we derive
\begin{eqnarray}
\label{loc}
    \hat{\kappa}_- (u,v)^T_{0,0,3} \,\, (\alpha - e^{-\lambda_{loc}}) 
    = 0
\end{eqnarray}
where it is taken into account that the coupling term with $1^{\text{st}}$ cite is zero because for our anzatz, $(u,v)^T_{1,1,1} = 0$. To satisfy Eq.~\eqref{loc}, either $(u,v)^T_{0,0,3}$ should be the eigenstate of $\hat{\kappa}_-$ with zero eigenvalue, or $\alpha = e^{-\lambda_{{\rm loc}}}$. The first option leads to the specific relation for the coupling constants $\Delta = \sqrt{\kappa \gamma}$, which includes the case of accidental flat minibands $\Delta = \kappa = \gamma$, but sets no bounds on the localization length, and thus does not correspond to the localized corner state. 
The latter option leads, for $\alpha < 1$, to the localization length \begin{eqnarray}
    \lambda_{{\rm loc}} = 1/|\ln{\alpha}|
\end{eqnarray} 
where we discard the nonphysical exponentially growing solutions. This result is similar to 1D SSH model~\cite{1DSSH}.

\section{Supplemental Note 6: Effect of imperfect mode degeneracy}

\begin{figure}[h!]
	\centering
    \includegraphics[width=0.6\textwidth]{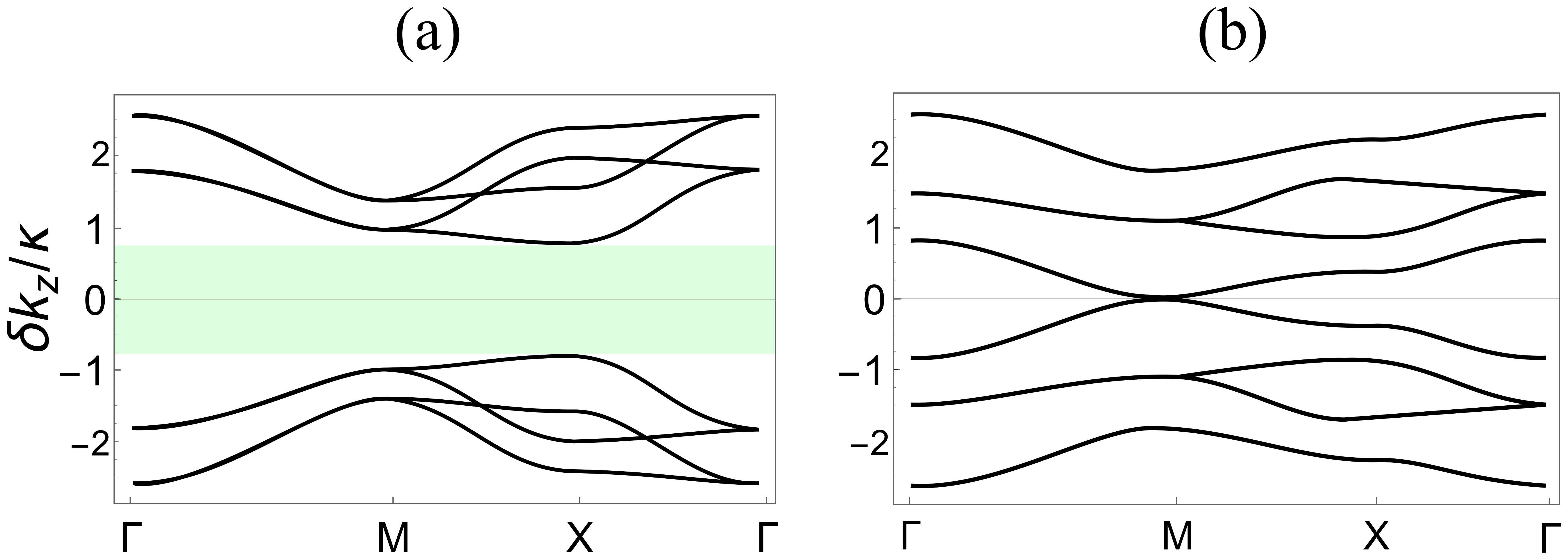}
    \caption{Band structure transformation with increasing wavevector detuning in each waveguide: (a) $\delta k = 0$, and (b) $\delta k = 2(1-\alpha)$ (parameters: $\kappa=\gamma$, $\Delta = 0.7 \kappa$, $\alpha = 0.3$). }
	\label{Fig_S_deltak}
\end{figure}

In this section, we discuss the effect of detuning of mode propagation constants caused by the imperfect waveguide design. For simplicity, we suppose that the individual waveguides are still identical, but the two modes of interest in each waveguide at given frequency have slightly different propagation constants, the difference being $ 2 \delta k = k_{z1} - k_{z_2} $. 
In the Hamiltonian Eq.~(2) of the main text, this is reflected in the appearance of diagonal terms, $\hat{H}'(\textbf{k}) = \hat{H}(\textbf{k}) + \delta \hat{H}$, where $\delta \hat{H} = (\delta k / 2) \cdot \hat{I}_4 \otimes \hat{\sigma}_z $, $ \hat{I}_4 $ is the four-dimensional unit matrix, and $\hat{\sigma}_z = \{1,0;0,-1\}$ is the corresponding Pauli matrix. 
Using this perturbed Hamiltonian, we plot the example of band structure in Fig.~\ref{Fig_S_deltak} and observe the topological gap closing at $M$ point for some critical value of wavevector detuning $\delta k_{c}$. 
This critical detuning therefore serves as a measure of maximum waveguide design imperfections which would not disturb the topological corner states. 

The spectrum of $\hat{H}'(\textbf{k})$ at $M$ point in the general case reads $\{-2 (\alpha -1) \gamma - \delta k, 2 (\alpha -1) \gamma - \delta k,-\sqrt{4 (\alpha -1)^2 \Delta ^2 + \delta k^2},-\sqrt{4 (\alpha -1)^2 \Delta ^2 + \delta k^2},\sqrt{4 (\alpha -1)^2 \Delta ^2 + \delta k^2},\sqrt{4 (\alpha -1)^2 \Delta ^2 + \delta k^2}, \delta k - 2 (\alpha -1) \kappa ,2 (\alpha -1) \kappa + \delta k\}$. The critical detuning $\delta k_{c}$ therefore corresponds to the intersection of at least one of the bands with zero energy, from what follows  
\begin{equation}
\delta k_{c} = 2 (1-\alpha ) \cdot \min( \gamma, \kappa ) 
. 
\end{equation}
This result is also reproduced by numerical calculations in finite lattices, where we find no corner state when $\delta k > \delta k_{c}$.



\section{Supplemental Note 7: Robustness of corner states to disorder}

\begin{figure}[h!]
	\centering
	\includegraphics[width=0.45\textwidth]{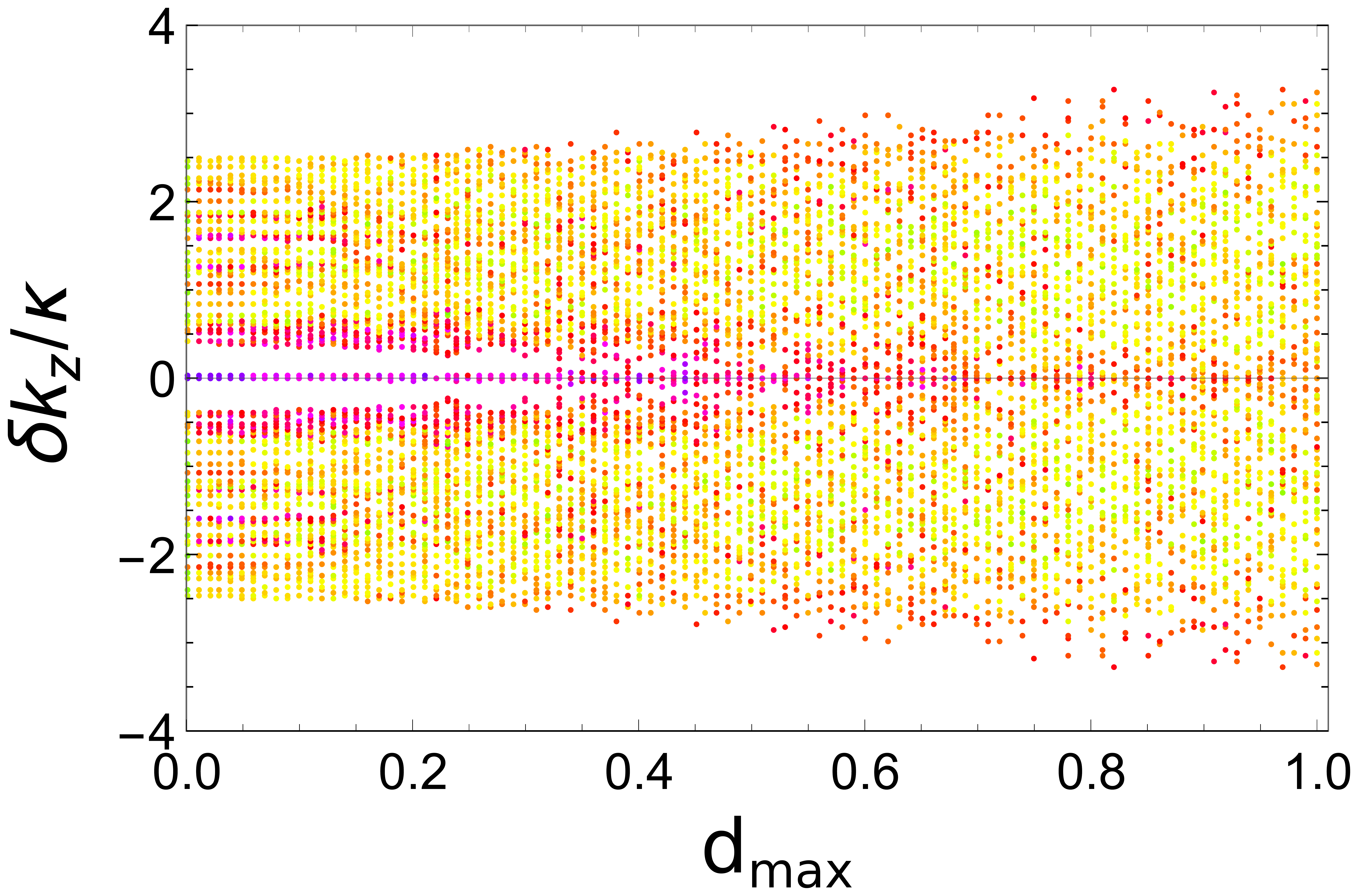}
	\caption{
		Evolution of the eigenstates of a $14\times 14$-site lattice when introducing the disorder. The horizontal axis sweeps the $d_{max}$ values. Logarithm of IPR in color coded as in the Fig.~3 in the main text. 
		Parameters: $\kappa=\gamma = 1$, $\Delta = 0.5$, $\alpha = 0.3$. 
	}
	\label{Fig_S_disorder}
\end{figure}

We introduce the disorder into the coupling matrices without breaking the hermiticity of the full tight-binding Hamiltonian. For each coupling matrix $\kappa_\pm$, we introduce three possible types of disorder: 
\begin{eqnarray}
&&\hat{M}_{d1}
=
d_1 \cdot 
\left(
\begin{array}{cccccc}
1 & 0 \\
0 & 0 \\
\end{array}
\right)
, \\
&&\hat{M}_{d2}
=
d_2 \cdot 
\left(
\begin{array}{cccccc}
0 & 0 \\
0 & 1 \\
\end{array}
\right)
, \\
&&\hat{M}_{d3}
=
d_3 \cdot 
\left(
\begin{array}{cccccc}
0 & i \\
-i & 0 \\
\end{array}
\right)
,
\end{eqnarray}
where $d_i \in [0,\kappa \cdot d_{max}],\, i \in \{1,2,3\}$ are three positive different disorder coefficients, and $d_{max}$ is a ``global'' maximum disorder constant normalized to $\kappa$ which is set for each consecutive eigenvalues calculation. 
The results for the finite $14\times 14$-site lattice are shown in Fig.~\ref{Fig_S_disorder}. We see that the zero-energy corner states survive and retain their near-zero energy until being connected via disorder to the bulk states at large enough $d_{max} \sim \kappa$. 
The stability of corner modes to moderate disorder confirms their topological origin, discussed in detail in the main text. 

\section{Supplemental Note 8: Realistic design at microwave frequencies}

\begin{figure}[h!]
	\centering
	\includegraphics[width=0.40\textwidth]{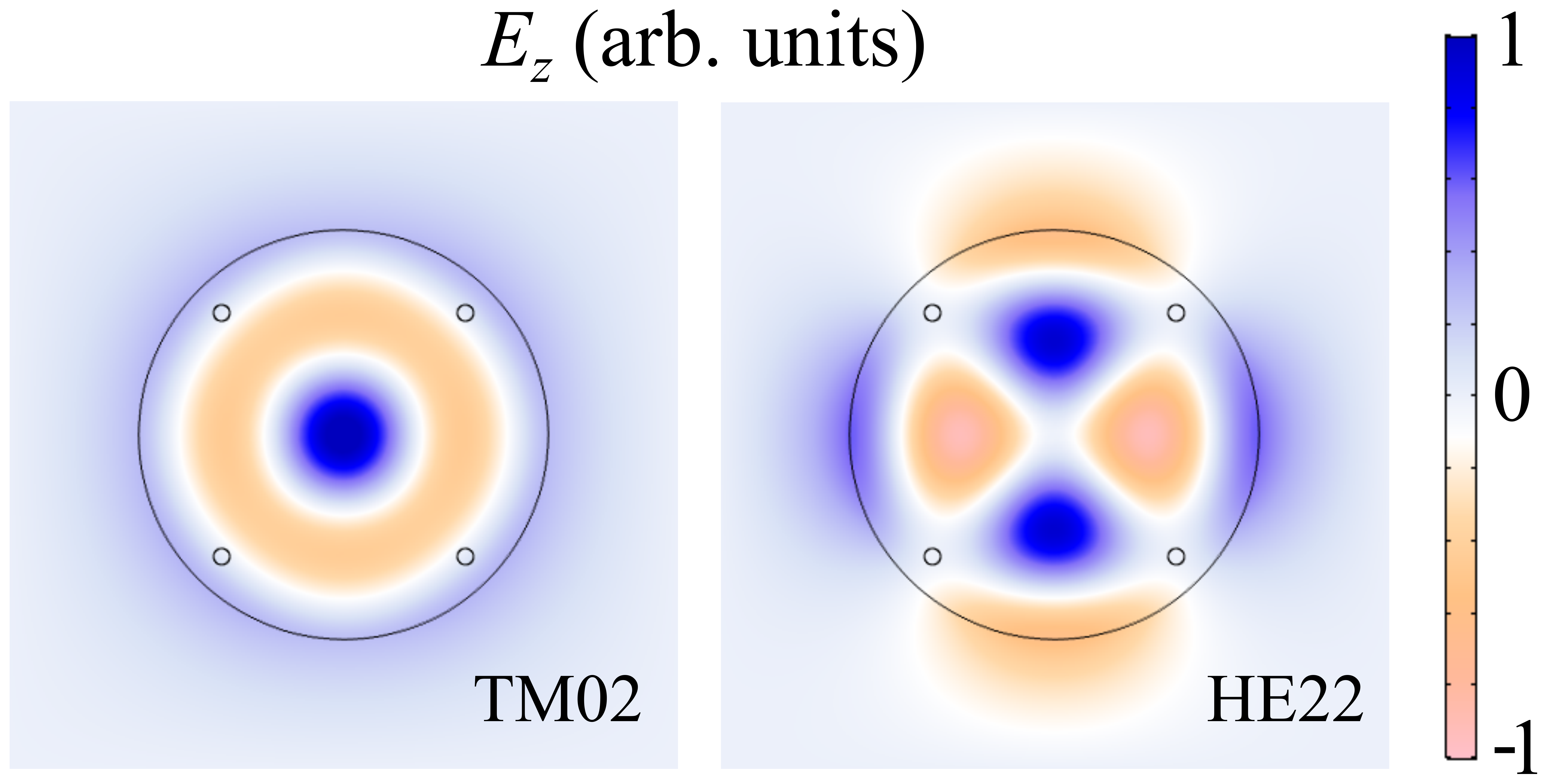}
	\caption{
		Two degenerate modes at frequency $f = 8.83 \,\text{GHz}$ for a waveguide described in the text. 
		The longitudinal component of the electric field is plotted. 
	}
	\label{degenerate_modes}
\end{figure}

In microwaves, we propose to exploit a simple waveguide of radius $1.26 \,\text{cm}$ and permittivity $\varepsilon = 9.9$ that corresponds to the commercially available ceramic aluminium oxide. The designed waveguide hosts the degenerate modes TM$_{02}$ and HE$_{22}$ shown in Fig.~\ref{degenerate_modes} at frequency $f = 8.83 \,\text{GHz}$, with propagation constants around $245 \,\text{rad/m}$ and degeneracy imperfection of the order of $1 \,\text{rad/m}$ (which is less than $0.5 \%$). 
The dominant electric field component responsible for the mode coupling in this case is the longitudinal component of electric field, $E_z$. 
Small circular holes are drilled out in order to spectrally separate the rotated partner HE$_{22}$ mode (which does not interact with two modes of interest in nearest-neighbour approximation) to minimize the intersection of the continuum band associated with this mode with the corner modes, while the degeneracy of the desired modes is retained. 
However, even without the drilled holes, it is in possible to access the topological corner states by the appropriate excitation, e.g. via auxiliary waveguide, or with a system of coherent dipoles.

\begin{figure}[h!]
	\centering
	\includegraphics[width=0.75\textwidth]{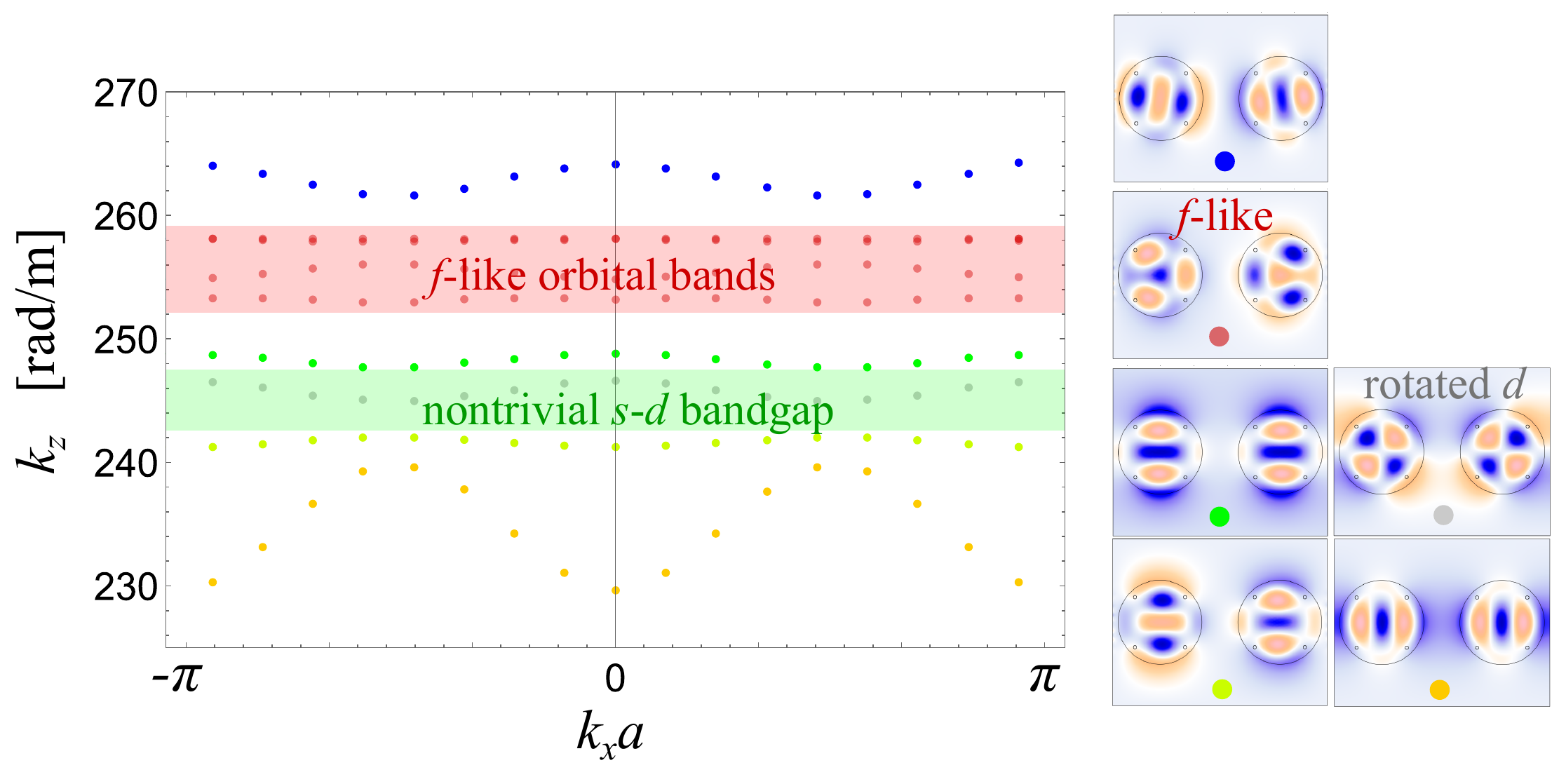}
	\caption{
		Band diagram of 1D SSH $s$-$d$ hybridized chain. This chain is identical to a row/column of the two-dimensional lattice presented in the main text. Panels on the right display corresponding eigenstates of a unit cell at $k_x = 0 $ to highlight distinct orbital types of the bands. 
	}
	\label{Fig_S_bands_fullwave_1D}
\end{figure}

\begin{figure}[H]
	\centering
	\includegraphics[width=0.55\textwidth]{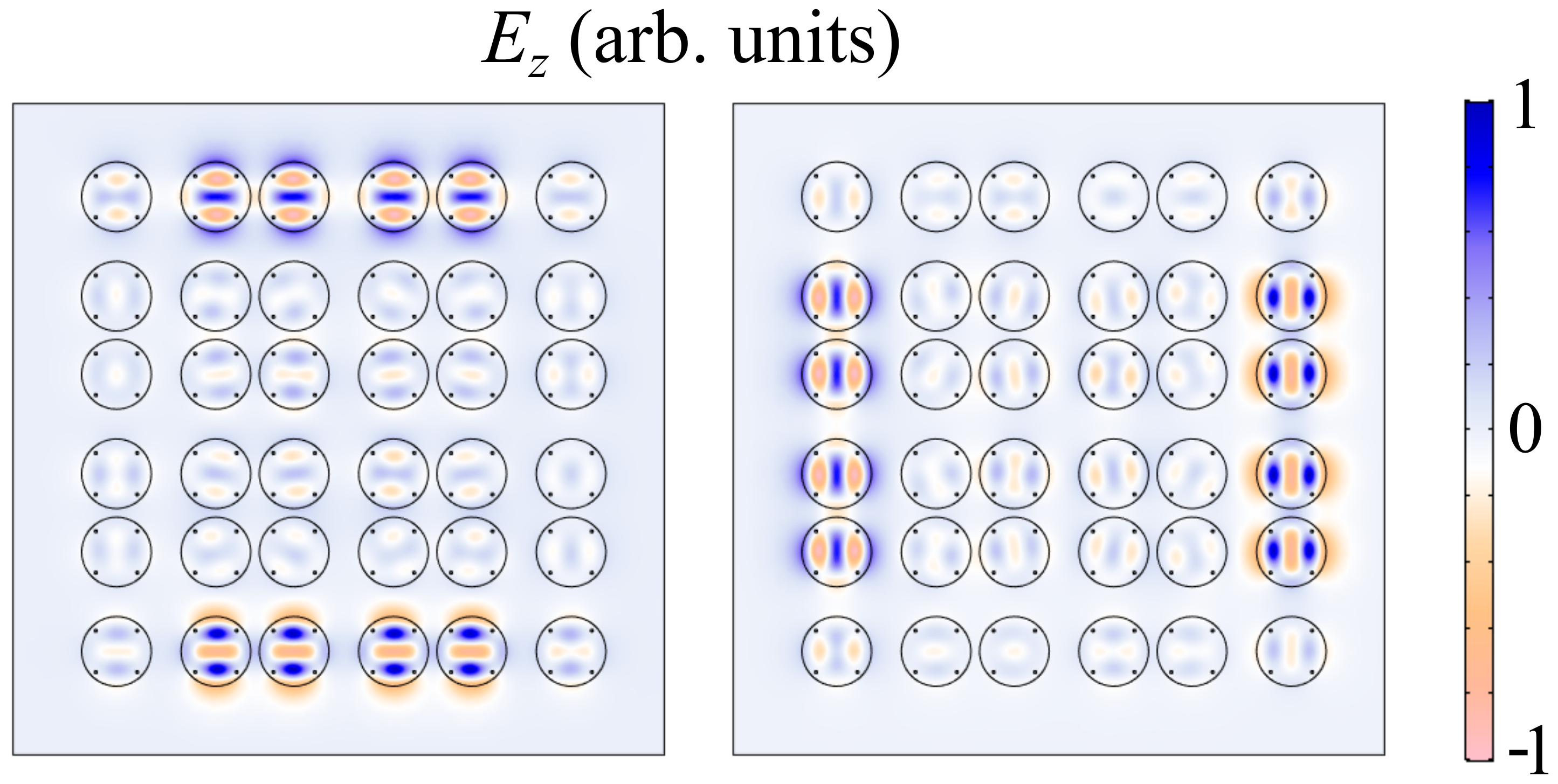}
	\caption{
		Typical edge states found at propagation constant $246.8 \,\text{rad/m}$, which consist of the $s$-$d$ hybridized on-site modes. The longitudinal component of the electric field is plotted. 
	}
	\label{Fig_S_edge_states_fullwave}
\end{figure}

Next, we calculate numerically and examine qualitatively the band diagram of the finite array. For simplicity, we analyze a one-dimensional array with the same inter-waveguide distances as in a row or a column of the two-dimensional lattice presented in the main text. 
The band diagram obtained from full-wave calculations with periodic boundary conditions is shown in Fig.~\ref{Fig_S_bands_fullwave_1D}. We observe that although the overall band structure with four $s$-$d$-hybridized bands is apparent, some additional orbital bands are also present. First, there are rotated $d$-orbital bands, one of which resides in the nontrivial bandgap of the $s$-$d$-hybridized bands. Second, $f$-like orbital states are present above the bandgap. 
Although coupling of $s$ and $d$ modes of interest to $f$-like orbital modes makes simplified analytical fit of the band structure insufficient, this does not preclude the correct prediction of the nontrivial $s$-$d$ bandgap (shown in green in Fig.~\ref{Fig_S_bands_fullwave_1D}) in the scope of simplified analytical model, as well as the observation of the corner states and the edge states (see Fig.~\ref{Fig_S_edge_states_fullwave}) in full-wave simulations. 
Thus, although the spectrum of the proposed waveguide lattice is more complex than that of the pure $s$-$d$ hybridized model, the latter proves to be sufficient to grasp correctly its higher-order topological properties.

\bibliography{refs}